\documentclass[10pt,conference]{IEEEtran}

\usepackage{xspace}
\usepackage{caption}
\usepackage{enumitem}
\usepackage{subcaption}
\usepackage{pifont}
\usepackage{tabu}
\usepackage{color}
\usepackage{booktabs}
\usepackage{cite}
\usepackage{amsmath,amssymb,amsfonts}
\usepackage{stmaryrd}
\usepackage{graphicx}
\usepackage{textcomp}
\usepackage{xcolor}
\usepackage{hyperref}

\newcommand{\cmark}{\ding{51}}%
\newcommand{\xmark}{\ding{55}}%

\usepackage{algorithm}
\usepackage[noend]{algpseudocode}

\makeatletter
\newenvironment{procedure}[1][htb]{%
    \renewcommand{\ALG@name}{Procedure}% Update algorithm name
   \begin{algorithm}[#1]%
  }{\end{algorithm}}
\makeatother

\usepackage{pgfplotstable}
\usepackage{pgfplots}
\definecolor{darkgreen}{rgb}{0,0.5,0.17}

\newcommand*{\plotscale}{0.80} % edit to scale plots
\renewcommand{\intextsep}{4pt}

\usepackage[colorinlistoftodos,prependcaption,textsize=tiny]{todonotes}

\usepackage{soul}
\usepackage[draft,inline,nomargin]{fixme}

\graphicspath{ {figures/} } % figures are located in ./figures

\newcommand{\tool}[1]{\textsc{#1}\xspace}

\newcommand{\clazz}[1]{\texttt{#1}\xspace}
\newcommand{\method}[1]{\texttt{#1}\xspace}
\newcommand{\prog}[1]{\mbox{\texttt{#1}}\xspace}
\newcommand{\openjdk}{\textsc{OpenJDK}\xspace}

\newcommand{\java}{\textsc{Java}\xspace}
\newcommand{\afl}{\tool{AFL}}

\newcommand{\kelinci}{\tool{Kelinci}}

\newcommand{\diffuzz}{\tool{DifFuzz}}
\newcommand{\themis}{\tool{Themis}}
\newcommand{\blazer}{\tool{Blazer}}

\usepackage{listings}

\usepackage{balance}
\IEEEoverridecommandlockouts

\begin{document}
\title{\diffuzz: Differential Fuzzing for Side-Channel Analysis}

%\author{\IEEEauthorblockN{Shirin Nilizadeh\IEEEauthorrefmark{1}}
\author{\IEEEauthorblockN{Shirin Nilizadeh\thanks{*Joint first authors}\IEEEauthorrefmark{1}}
\IEEEauthorblockA{\textit{University of Texas at Arlington} \\
Arlington, TX, USA \\
shirin.nilizadeh@uta.edu}
\and
\IEEEauthorblockN{Yannic Noller\IEEEauthorrefmark{1}}
\IEEEauthorblockA{\textit{Humboldt-Universit{\"a}t zu Berlin} \\
Berlin, Germany \\
yannic.noller@hu-berlin.de}
\and
\IEEEauthorblockN{Corina S. P\u{a}s\u{a}reanu}
\IEEEauthorblockA{\textit{Carnegie Mellon University Silicon Valley,} \\
\textit{NASA Ames Research Center} \\
Moffett Field, CA, USA \\
corina.s.pasareanu@nasa.gov}
}

\maketitle

\begin{abstract}
Side-channel attacks allow an adversary to uncover secret program data by observing the behavior of a program with respect to a resource, such as execution time, consumed memory or response size. Side-channel vulnerabilities are difficult to reason about as they involve analyzing the correlations between resource usage over multiple program paths.
We present \diffuzz, a fuzzing-based approach for detecting side-channel vulnerabilities related to time and space. \diffuzz automatically detects these vulnerabilities by analyzing two versions of the program and using 
resource-guided heuristics to find inputs that {\em maximize} the difference in resource consumption between secret-dependent paths.
The methodology of \diffuzz is general and can be applied to programs written in any language.
For this paper, we present an implementation that targets analysis of \java programs, and uses and extends the \kelinci and \afl fuzzers. 
We evaluate \diffuzz on a large number of \java programs and demonstrate that it can reveal unknown side-channel vulnerabilities in popular applications.
We also show that \diffuzz compares favorably against \blazer and \themis, two state-of-the-art analysis tools for finding side-channels in \java programs. 
%\diffuzz found new vulnerabilities in repaired programs that were shown to be safe with these previous tools.
\end{abstract}

%\begin{CCSXML}
%<ccs2012>
%<concept>
%<concept_id>10002978.10003022.10003023</concept_id>
%<concept_desc>Security and privacy~Software security engineering</concept_desc>
%<concept_significance>500</concept_sigtnificance>
%</concept>
%<concept>
%<concept_id>10002978.10002986.10002990</concept_id>
%<concept_desc>Security and privacy~Logic and verification</concept_desc>
%<concept_significance>100</concept_significance>
%</concept>
%</ccs2012>
%\end{CCSXML}

%\ccsdesc{Security and privacy~Use https://dl.acm.org/ccs.cfm to generate actual concepts section for your paper}
% -- end of section to replace with generated code


%\footnote{\IEEEauthorrefmark{1}{Joint first authors}}
\section{Introduction}
%The widespread use of computing systems in everyday life has increased access to %sensitive data, ranging from personal information of individual users to trade and %military secrets of states and corporations.
%There is thus an increasing need for techniques and tools that can ensure that computing %systems manipulate sensitive data in a secure manner.
%This is particularly hard to achieve in the face of 
Side-channel attacks enable an adversary to uncover sensitive information from programs by observing non-functional characteristics of program behavior, 
such as execution time, memory usage, response size, network traffic, or power consumption. 
There is a large literature on side channels showing evidence that they are practical and can have 
serious security consequences~\cite{Brumley:2003:RTA:1251353.1251354, cachepractical, ASLRtiming}.
For instance, exploitable timing channel information flows were discovered for  Google's Keyczar Library~\cite{Law09.2}, the Xbox 360~\cite{Xbox} and implementations of RSA encryption~\cite{Brumley:2003:RTA:1251353.1251354}.
%not longer available: (yn), and the open authorization protocol OAuth~\cite{OAuth13}.
More recently, the Meltdown and Spectre side-channel attacks~\cite{meltdown} have shown how to exploit critical vulnerabilities in modern processors to uncover secret information.
These vulnerabilities highlight the increased need for tools and techniques that can effectively discover side channels before they are exploited by a malicious user in the field. However, side-channel vulnerabilities are difficult to reason about as they involve analyzing correlations between resource usage over multiple program paths.

In this paper we present \diffuzz, a dynamic analysis approach for the detection of side channels in software systems. 
Given a program whose inputs are partitioned into public and secret variables, \diffuzz uses a form of differential fuzzing to automatically find program inputs that reveal side channels related to a specified resource, such as time, consumed memory, or response size.
%In this work
We focus specifically on timing and space related vulnerabilities, but the approach can be adapted to other types of side channels, including cache based.

Differential fuzzing has been successfully applied before for finding bugs and vulnerabilities in a variety of applications, such as LF and XZ parsers, PDF viewers, SSL/TLS libraries, and C compilers \cite{DBLP:conf/sp/PetsiosTSKJ17, Sivakorn2017, Yang2011}.
However, to the best of our knowledge, we are the first to explore differential fuzzing for side-channel analysis. 
Typically such fuzzing techniques analyze different versions of a program, attempting to discover bugs by observing differences in execution for the same inputs.
In contrast \diffuzz works by analyzing two copies of the same program, with the same public inputs but with different secret values, and computing the {\em difference} in side channel measurements (time or space) observed over the two executions.
If the difference is large, then it means that the program has a side-channel vulnerability, which should be 
remedied by  the developer.

The approach is similar to the well-known method of {\em self-composition}~\cite{Barthe04}), which is used to check that no matter what the secret is, the program yields the same output.
%\yannic{@Corina: it does not need to be the same output, but the same observable cost, right? Classic self composition checks the output}
If that is the case, the program is said to satisfy {\em non-interference}, meaning that the program leaks {\em no} information; otherwise, the program is vulnerable. However, it has been argued~\cite{themis,blazer} that non-interference with regard to side channels is too strong a property for most realistic programs, as it is almost always the case that some variation in resource usage, particularly execution time, exists for different program paths. If the difference is small, it may not be exploitable in practice, since it may not be actually observable by an attacker. For example, consider the case of a client-server application. Small variations in execution time on the server side may not be observable (and therefore exploitable) on the client side.
In such cases the program can be considered secure although it does not satisfy non-interference. If on the other hand, the difference is large, this indicates a side-channel vulnerability since an attacker can use differences between measurements to distinguish between secrets. For this reason, \diffuzz does not merely check non-interference, but instead employs resource-guided heuristics to automatically find inputs that attempt to {\em maximize} the difference in resource consumption between secret-dependent paths.

%To detect side-channels, \diffuzz uses resource-guided heuristics to automatically find (secret and public) inputs that {\em maximize} %the difference $\delta$ in resource consumption between secret-dependent paths. If the $\delta$ is found to be zero this corresponds to %the well-known notion of {\em non-interference}, meaning that no vulnerability could be found. If on the other hand, the computed $\delta%$ is greater than zero, this means that the program has a vulnerability. As not all the differences in resource usage are exploitable in %practice, since they might be too small to be actually observable by an attacker, we also specify a threshold $\epsilon$ and report a %side-channel vulnerability only if the computed $\delta$ exceeds this threshold. 

% actually this figure appears in the next section but I would like to have on the second page of the paper
\begin{figure*}
\centering \includegraphics[width=.95\linewidth]{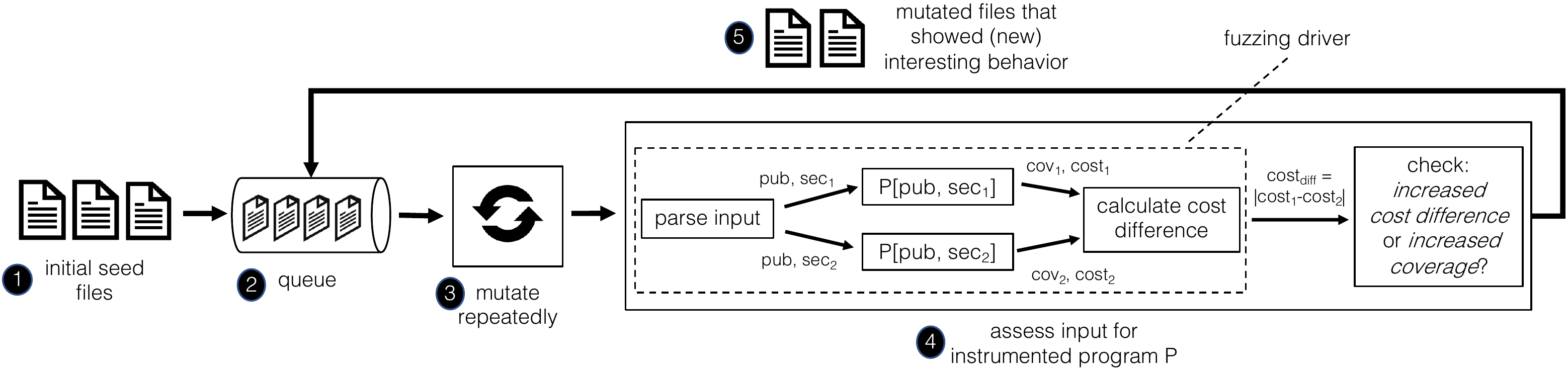}
\caption{Overview of \diffuzz approach.}
\label{fig:overview_approach}
\end{figure*}

The methodology that we advocate with \diffuzz is general and can be applied to programs written in any language.
%implemented in any fuzzer, with some modifications. 
For this paper we present an implementation that targets \java programs and is based on \tool{American Fuzzy Lop} (\afl) \cite{afl} and \kelinci~\cite{kelinci_ccs2017}. \afl is a fuzz testing tool that uses genetic algorithms to mutate user-provided inputs using byte-level operations with the goal of increasing coverage; \kelinci provides an interface to execute \afl on \java programs. 
To perform side-channel analysis, \diffuzz instruments a program to record resource consumption, in addition to coverage, along the paths that are executed by the fuzzed inputs.
%To perform side-channel analysis, \diffuzz measures the time and space consumption along the paths executed by the fuzzed inputs.
%computes an approximate path cost, in terms of the number of instructions executed along a path. To discover space side-channels, \diffuzz measures memory %and space consumption, by frequently polling the current memory used in the Java Virtual Machine and measuring the size (in bytes) of the values that are %returned and the messages that are sent by the program.
%as well as the memory and space consumption, for space side-channels.
%\yannic{! this is not correct: the instrumentation is only for timing; the space information is measured by frequently polling the current memory used %in the Java Virtual Machine.}
Furthermore, \diffuzz records the {\em difference} in resource consumption in a user-defined cost. This difference is sent back to the fuzzer, whose mutants are marked as {\em important} if there is an increase in the computed difference, thus {\em guiding} the fuzzer towards inputs that expose vulnerabilities.

We have applied \diffuzz on well-known, widely used \java applications, such as Apache FtpServer~\cite{ftpserver} and AuthMeReloaded~\cite{AuthMeReloaded}, where we found new, previously unknown, vulnerabilities, which were confirmed by the developers. 
Additionally we have applied our approach on complex examples from the DARPA Space/Time Analysis for Cybersecurity (STAC) program~\cite{stac}, IBASys, an image-based authentication system, and CRIME, an instance of the Compression Ratio Info-leak Made Easy attack~\cite{CRIMEattack}, where we found vulnerabilities related to both time and space consumption. %{\bf Shirin: please add the new examples}

We also compared \diffuzz with \blazer \cite{blazer} and \themis \cite{themis}, two state-of-the-art analysis tools for finding side channels in \java programs. 
%Both tools can analyze side channels by checking a form of non-interference with respect to a bound $\epsilon$. 
Both tools perform static analysis and can in principle guarantee absence of side channels, but may also give false alarms due to underlying over-approximation.
In contrast \diffuzz performs a dynamic analysis, and thus does not give false alarms (provided that the fuzzing driver is meaningful, see Section 2.3.), but it can not prove absence of vulnerabilities.

%In practice, both \blazer and \themis are {\em incomplete}, as they can not handle important features of \java, such as recursion, %reflective calls, dynamic class loading and exception handling (see Section 7 in the \themis paper \cite{themis}) and can not cope with %external libraries, which need to be abstracted leading to more imprecision.
%Scalability is also an issue; for this reason these tools perform some pre-processing to focus only on some important methods.
%In contrast, \diffuzz can handle all \java programs and can in principle scale to arbitrary sizes, although in that case it might be very %slow to converge to the maximum difference. We address the problem with resource-guided heuristics.
%Furthermore, \diffuzz generates actual test inputs that expose the vulnerabilities and can be used by developers to debug and fix the %code.

We evaluated \diffuzz on the same benchmarks from \themis and \blazer and were able to find the same vulnerabilities.
We also ran \diffuzz on the corrected (safe) versions (when they were available).
For the majority of these cases, we found that as expected, \diffuzz correctly finds zero or a small differences %while at the same time showing good code coverage, 
thus showing the usefulness of the approach also in the case of absence of vulnerabilities.
However, we have also found that, in some cases, \diffuzz uncovered new vulnerabilities in versions which were shown to be safe with \blazer and \themis.
%This is due to some simplifying, unstated assumptions made in the previous tools. For example, they use some manually constructed models %that are inaccurate, they do not handle overflow or they operate at the level of the Jimple intermediate representation~\cite{SOOT}, %where some operations are constant time whereas at bytecode level they are not. These findings highlight the subtlety of side-channel %vulnerabilities and the importance of developing technique and tools that can effectively find them.

%The \blazer and \themis benchmarks consist of small examples of side-channel vulnerabilities from the literature as well as real world %applications with \emph{known} timing and space side-channels.
%Additionally we apply our approach on two examples from the DARPA Space/Time Analysis for Cybersecurity (STAC) program~\cite{stac}, %IBASys, a complex image-based authentication system, and CRIME, an instance of the CRIME (Compression Ratio Info-leak Made Easy) attack~%%\cite{CRIMEattack}.
%To demonstrate that our approach can find unknown side-channel vulnerabilities in well-known, widely used \java applications, we applied %\diffuzz on Apache FtpServer~\cite{ftpserver} and AuthMeReloaded~\cite{AuthMeReloaded}, where we also found new vulnerabilities, which %were confirmed by the developers. 

In summary, this work makes the following contributions:
\begin{itemize}
\item We present \diffuzz, the first differential fuzzing approach for finding side-channel vulnerabilities.
\item We evaluate \diffuzz on multiple security-critical \java applications and we report new vulnerabilities in well known \java applications, such as Apache FtpServer.
\item We compare with state-of-the-art tools \blazer and \themis, where we highlight some new vulnerabilities in programs that were previously deemed safe.
\end{itemize}

\section{Approach}
\label{sec:approach}

%\begin{figure*}
%\centering \includegraphics[width=.95\linewidth]{overview.pdf}
%\caption{Overview of \diffuzz approach.}
%\label{fig:overview_approach}
%\end{figure*}

%\yannic{@Corina: added driver to picture and mentioned instrumentation in caption. Is this okay?}

Figure \ref{fig:overview_approach} shows the overview of our differential fuzzing approach.
%Step 1, 2, 3 and 5 are similar to the general mutation-based fuzzing approach, e.g., like implemented in \afl \cite{afl}, %which we inherit by building our approach based on \kelinci \cite{kelinci_ccs2017}.
%Step 4 represents our new input assessment strategy that allows \diffuzz to reuse existent cost-guided fuzzing approaches for %the efficient side-channel analysis.
To start the analysis, the user needs to provide initial seed files that exercise the program under test (cf. step 1 in Figure \ref{fig:overview_approach}). The user also needs to provide a {\em driver}, which {\em parses} an input file into three elements $pub$ (common public value), $sec_1$, and $sec_2$ (two secret values, one for each program copy) and executes two copies of the program on these inputs.
The program is {\em instrumented} to record resource consumption and coverage information.

The seed files are put into a queue for further processing (cf. step 2 in Figure \ref{fig:overview_approach}).
This queue is used during the whole process as the central data structure that includes all the inputs that are deemed \emph{interesting} by the analysis.
The fuzzer will take the inputs from the queue and will mutate them repeatedly (cf. step 3 in Figure \ref{fig:overview_approach}).
In order to decide whether a mutated input is interesting for further processing, \diffuzz executes the driver with this input, computes the cost difference between two executions, which is handled as the score for this input, and compares it with the maximum cost difference (aka cost difference \emph{high-score}), which was observed in the previous executions (cf. step 4 in Figure \ref{fig:overview_approach}).
%Additionally, the Figure \ref{fig:overview_approach} shows in step 4 which part of the assessment is implemented in the fuzzing driver (cf. %sections below).

Only the inputs that either lead to increased high-score or to increased overall program coverage will be forwarded to the fuzzing queue (cf. step 5 in Figure \ref{fig:overview_approach}).
The process is repeated until a user-specified timeout occurs. 
%\diffuzz output consists of concrete input files that trigger the obtained cost differences, including the maximum difference, and the %achieved coverage information.

%In a recent work B{\"o}hme \cite{stads} presented a statistical framework to assess the residual risk that a vulnerability exists when no vulnerability %was discovered with fuzzing until the user-specified timeout. 
%In future, we plan to augment our side-channel analysis with such a framework to further support the user and strengthen the confidence in the obtained %results.

We describe the \diffuzz approach in more detail below.

\subsection{Side-Channel Analysis}\label{subsec:side-channel_analysis}
Information flow analysis is typically used to determine that a program manipulates secret data in a secure manner.
The analysis accepts programs as secure if the secret data can not be inferred by an attacker through their observations of the systems.
This intuitive property is called {\em non-interference}. In the case of side-channels, the {\em observations} consist of the side-channel measurements that an attacker can make.

There are many techniques for checking non-interference.
The simplest one is through {\em self-composition}~\cite{Barthe04}.
At a high level the technique reduces the problem of secure information flow of a program to analyzing two copies of the same program, where the secret inputs are renamed, but the public values stay the same, and checking that these two copies create the same observation. 

Let $P$ be a program, and $P\llbracket pub,sec \rrbracket$ be the execution of the program $P$ with inputs $pub$ and $sec$.
As it is customary in the security literature, we break down the program inputs to a tuple of public (low) values and secret (high) values.
We abbreviate the public values as $pub$ and the secret values as $sec$.
Furthermore let $c(.)$ be the evaluation of a program execution with respect to a particular cost encoding the resource usage (e.g., execution time or response size) of the program.
The non-interference requirement can then be formalized as follows:
$$\forall pub, sec_1, sec_2: c(P\llbracket pub, sec_1 \rrbracket) = c(P\llbracket pub, sec_2 \rrbracket)$$

Intuitively, the property states that any two secrets are {\em indistinguishable} through the side-channel observations and therefore can not be uncovered by an attacker.

Although satisfying non-interference is a sound guarantee for a system to be secure, this requirement is too strict for the side-channel analysis of most realistic programs.
Particularly for timing channels, small differences in computations may be imperceptible to an attacker and can thus not be exploited in practice.
This problem was observed in various papers before~\cite{blazer, themis} and was formalized as checking $\epsilon$-bounded non-interference in~\cite{themis}: not only programs with zero interference can be accepted as secure, but also programs where the difference between observations is too small (below a threshold $\epsilon$) to be exploitable in practice.
Thus the program is deemed to be secure if the following condition holds:
$$\forall pub, sec_1, sec_2: | c(P\llbracket pub, sec_1 \rrbracket) - c(P\llbracket pub, sec_2 \rrbracket)| < \epsilon$$

One can perform the above check by enumerating all the possible input combinations, measuring the resource consumption for each run, and performing the check for the two versions of the program, but this could become quickly intractable for most realistic programs.
%Existing software model checkers~\cite{Visser2003, VeriSoft, Spin} have shown promise in solving such combinatorial problems, however they are not %equipped to check non-functional properties related to resource usage and therefore can not be readily used to check for side-channels. Furthermore, %analyzing two copies of the programs is likely infeasible in practice due to the large number of paths that need to be analyzed.

We therefore advocate the use of fuzzing to address the problem. However typical fuzzing tools are engineered to only increase code coverage and can thus be very slow in generating inputs that expose a significant difference in resource consumption. The key ingredient of our approach is the incorporation of heuristics that guide the fuzzing towards configurations that maximize this difference, as explained in the following sections. Note that, unlike previous techniques, that use static analysis to check $\epsilon$-bounded non-interference~\cite{blazer,themis}, we do not require the user to provide an a-priori threshold $\epsilon$; instead we let the tool try to maximize the difference between secret-dependent paths.

\subsection{Attacker Model}
%Before describing our technique, 
We review here the attacker model considered in this paper, which is similar to previous work on the topic~\cite{themis,blazer}.
We assume the program is deterministic and that the side-channel measurements are precise.
We further assume that the attacker can not observe anything else (i.e., the attacker does not use the main-channel to infer information). 
When measuring resource usage we assume that any variations are caused by the application software, and we are thus ignoring side-channels related to the hardware architecture or the physical environment.
In principle we can handle all these side-channels by using an available model of the corresponding resource.
Even in the absence of a model, we could use the inputs generated by the fuzzer to run the programs on a specific platform and perform actual, precise measurements with respect to the resource of interest. Furthermore, we could measure the wall-clock time and also the JIT (just-in-time compilation) effect. 
%In our experiments with timing side-channels, we used instruction counts, which provides a good approximation, allowing us to also compare with %previous approaches.

The mentioned assumptions are realistic. For example imagine a server-client scenario in a distributed environment (similar described in \cite{themis}), in which the attacker is physically separated from the victim application, i.e. there is no chance to observe any physical side-channel.
For an encrypted network communication the attacker cannot read the content of the sent messages, and hence, relies on the metrics that can be observed during communication with the server, like response sizes and response times.
Additionally, based on the physical distribution the attacker should not have the possibility to manipulate the victim application to observe any hardware-level side-channels.

Note that \diffuzz is also applicable to non-deterministic code and in the experiments we report on such an application. However, in general the results could be imprecise in this case, due to the {\em noise} introduced in the measurements. More analysis would be necessary, which is left for future work.

\subsection{Differential Fuzzing}\label{subsec:difFuzz}
Our approach aims to use fuzzing to analyze the two copies of the program and to {\em guide} it to find inputs that maximize the cost difference between two program executions, for which only the secret values are different:
\begin{equation} \label{eq:methodology}
\underset{pub,sec_1, sec_2}{\text{maximize:}} \delta = |c(P\llbracket pub, sec_1 \rrbracket) - c(P\llbracket pub, sec_2 \rrbracket)|
\end{equation}

\subsubsection*{{\bf Fuzzing Driver}}
In order to apply fuzzing we need a {\em driver} that parses the inputs from the fuzzer and executes the 
two copies of the code under test, while also measuring the cost difference.
Procedure~\ref{alg:diffuzzdriver} shows the general driver of our fuzzing approach.
It starts with parsing the input (cf. line 1), i.e., reading three different input values: the public value and two secret values, which are used to execute the program twice, as formulated in Equation \ref{eq:methodology}.
Additionally, the parsing can take some simple constraints for these input values, as described below.
For each execution we measure the costs (cf. line 2 and 3) and calculate the absolute cost difference (cf. line 4).
This value is used to guide the fuzzer (cf. line 5) to generate more inputs with the goal of increasing the difference. 
%{\em Cost-guided} fuzzers (cf. Section \ref{sec:fuzzing-background}) can be adapted for side-channel analysis as we %describe here, provided that they can be made aware of cost difference.
In our implementation this notion of cost difference is realized by setting user-defined cost values.

\begin{procedure}
\caption{Differential Fuzzing Driver}\label{alg:diffuzzdriver}
\begin{algorithmic}[1]
\State $\textit{pub}, sec_1, sec_2 \gets \text{parse(}input, constraints\text{)}$
\State $cost_1 \gets \text{measure($P$($pub, sec_1$))}$
\State $cost_2 \gets \text{measure($P$($pub, sec_2$))}$
\State $cost_{diff} \gets |cost_1 - cost_2|$
\State $\text{setUserDefinedCost}(cost_{diff})$
\end{algorithmic}
\end{procedure}

\subsubsection*{{\bf Input Constraints}}
Solving the maximization problem describ\-ed in Equation \ref{eq:methodology} for two totally arbitrary chosen input tuples might not be expedient because most applications assume certain properties of the secret values.
For example if a password is stored as a hash, the application would assume that the hashed values have the same fixed length. 
%(which would be a valid assumption).
Using secret values with arbitrary lengths for testing this application would lead to results that are not meaningful.
Therefore, in practice it is useful to set some simple constraints on the inputs.
In our approach we have a constructive solution, i.e. we rely on the user to encode input constraints in the driver such that only the inputs that satisfy these constraints are passed to the programs.
%our fuzzing driver derives only inputs that fulfill these properties.
For example the driver can limit the size of a string during parsing by simply not reading more characters than a given threshold, or the driver can ensure a certain character set for a string that should represent a hash value by mapping all non-member characters to member characters during parsing.

\subsubsection*{{\bf Analysis Outcome}}
The result of the analysis is a set of concrete public and secret inputs that expose the maximum cost difference between two secret-dependent paths found by the fuzzer.
If the difference is large, it indicates a side-channel vulnerability and the developer can use the provided inputs to precisely pinpoint the problem and fix the vulnerability, e.g., by making the cost similar on both program paths.
If on the other hand the difference is small (or zero) it could mean that the program has no vulnerabilities or that the fuzzer was not run long enough.
The developer can then run the fuzzer longer to get enough confidence that the software indeed has no vulnerability.
The fuzzer also records the {\em coverage} achieved on the analyzed code and this information can also be examined to increase the confidence in the reported results.

%In a recent work B{\"o}hme \cite{stads} presented a statistical framework to assess the residual risk that a vulnerability exists when no vulnerability was %discovered with fuzzing until the user-specified timeout. 
%In the future, we plan to augment our side-channel analysis with such a framework to further support the user and strengthen the confidence in the obtained %results.

\subsubsection*{{\bf Manual Effort}}
\diffuzz requires manual effort in writing the drivers and the input constraints. 
In the driver, the user needs to specify: how to parse the input file to retrieve valid input values, the entry point to start the target application, and  how to measure the execution cost. As many applications come with test cases, we use them to determine entry points. 
We believe that the manual effort is not high as all the drivers are very similar (and follow Procedure~\ref{alg:diffuzzdriver}); the constraints are minimal and application specific (e.g., passwords have certain lengths). One can also envision using fuzzing to discover these constraints automatically, following the work on grammar inference from~\cite{zellerASE16}. 
However this is left for future work.

%The drivers can be complex in terms of how to parse data: too many constraints and assumptions on the input data can narrow down the actual search space %while too loose constraints can lead to meaningless results in the analysis. Additionally, it requires some knowledge about the target application in order %to specify good entry points.

\subsection{Fuzzing Programs}
\label{sec:fuzzing-background}

For fuzzing we use off-the-shelf tools such as \afl \cite{afl}. %We use an AFL-based fuzzing for Java programs.
\afl is a state-of-the-art, security-oriented grey-box fuzzer that employs compile-time instrumentation and genetic algorithms to automatically generate test inputs that improve the branch coverage of the analyzed code.
%It supports programs written in C, C++, or Objective C, compiled with either \tool{GCC} or \tool{Clang}.
%In order to make it applicable to \java programs, we use \kelinci~\cite{kelinci_ccs2017}, which provides 

Fuzz testing tools have been very successful at finding bugs and vulnerabilities in a variety of applications, ranging from image processors and web browsers to system libraries and various language interpreters.
For example, \afl was instrumental in finding several of the Stagefright vulnerabilities in Android, the Shellshock related vulnerabilities, %CVE-2014-6277 and CVE-2014-6278, 
vulnerabilities in \tool{BIND}, %(CVE-2015-5722 and CVE-2015-5477), 
as well as numerous bugs in (security-critical) 
applications and libraries such as \tool{OpenSSL}, \tool{OpenSSH}, \tool{GnuTLS}, \tool{GnuPG}, \tool{PHP}, \tool{Apache}, \tool{IJG jpeg}, \tool{libjpeg-turbo} and many more (cf. bug list on AFL's website \cite{afl}).
Motivated by the success of fuzzing, we aim to use this technology for finding side-channel vulnerabilities. Typically, fuzzers use heuristic algorithms to mutate user-provided inputs to increase coverage, with the goal of finding crashes and other vulnerabilities. In contrast, \diffuzz uses fuzzing to perform a relational analysis, where the goal is to maximize the difference in resource usage for two copies of the program.

To realize this goal, an off-the-shelf fuzzer can be extended as follows: (1) the instrumentation is modified to collect additional information related to a resource consumption, such as timing, memory usage and response size; and (2) the {\em difference} between the costs observed for two program copies is recorded and sent back to the fuzzer, whose logic is modified to consider as {\em important} the inputs that increase this difference. In particular, the fuzzer maintains the so far observed difference \emph{high-score} and prioritizes inputs leading to new high-score in addition to improved coverage, attempting to maximize the difference and thus find side-channels.

The timing cost is approximated by counting every (bytecode) instruction executed by the program. A similar cost is used in previous static analysis tools (\themis and \blazer) allowing us to compare with them. 
Note that we can also measure the wall-clock time directly, by recording the execution time for each execution. The measurements can be performed on a clean, un-instrumented version of the program, using the inputs provided by fuzzing.
However, we found that these measurements could sometimes be imprecise due to garbage collection and other processes running on the same machine. One can perform multiple runs for the same input, and take the average of these measurements, but we did not explore this direction further in this work, as we found that counting the instructions provides a good approximation. 

Memory usage is measured by intermittent polling using a timer, which results in measuring the maximum consumption at any point during program execution. \diffuzz also measures response size (in bytes) for the values that are returned and the messages that are sent by the application.

%Furthermore, a user-defined cost is specified by the developer directly with an extra call to the method %\method{Kelinci.addCost(long)}.
%We use this user-defined cost to encode the difference in cost for two program executions and instruct the fuzzer to find inputs %that maximize it. We note that recent work~\cite{badger} also extends Kelinci with the ability of measuring costs; however, the %goal there is to perform a different kind of analysis, namely to estimate the worst-case complexity of programs.

We note that \afl supports programs written in C, C++, or Objective C.
%, compiled with either \tool{GCC} or \tool{Clang}.
To make it applicable to \java programs, we use \kelinci~\cite{kelinci_ccs2017}, which provides an \afl-style instrumentation for \java programs, executes the instrumented programs and sends results back to a simple C program that interfaces with  \afl. % (cf. Figure \ref{fig:kelinci}).
\afl does not know about the \java program in the background because it only communicates with the mentioned interface program, and hence, \afl can use its  heuristics to generate inputs that are then executed on the Java programs.

\subsection{Example}
We illustrate the side-channel analysis on a password checking example.
Listing~\ref{lst:unsafe-pwcheck} and Listing~\ref{lst:safe-pwcheck} show the code for the comparison of a user password with a server-side stored password in an \textit{unsafe} and \textit{safe} way respectively.

\begin{lstlisting}[caption=Unsafe Password Checking, label=lst:unsafe-pwcheck, numbers=left,  stepnumber=1, xleftmargin=1em]
boolean pwcheck_unsafe(byte[] pub, byte[] sec) {
	if (pub.length != sec.length) {
		return false;
	}
	for (int i = 0; i < pub.length; i++) {
		if (pub[i] != sec[i]) {
			return false;
		}
	}
	return true;
}
\end{lstlisting}

\begin{lstlisting}[caption=Safe Password Checking, label=lst:safe-pwcheck, numbers=left,  stepnumber=1, xleftmargin=1em]
boolean pwcheck_safe(byte[] pub, byte[] sec) {
	boolean unused;
	boolean matches = true;
	for (int i = 0; i < pub.length; i++) {
		if (i < sec.length) {
			if (pub[i] != sec[i]) {
				matches = false;
			} else {
				unused = true;
			}
		} else {
			unused = false;
			unused = true;
		}
	}
	return matches;
}
\end{lstlisting}
The \textit{unsafe} variant contains a timing side channel because its early-return in lines 2 and 6.
These two locations were fixed in the \textit{safe} variant by iterating over the complete password, even when the two passwords already do not match at an earlier point.
To apply \diffuzz, we built a driver for the \textit{unsafe} variant based on Procedure \ref{alg:diffuzzdriver} with length limit of 16 bytes for each fuzzed value (see Listing \ref{lst:pwcheck-driver}).
The way we parse the input in this example ensures that all three values have same length.
The field \prog{Mem.instrCost} holds the current cost measured by the instrumentation, i.e., in our case the number of executed bytecode instructions.
The method \prog{Mem.clear()} resets the current cost, which is necessary to measure the cost for each execution separately.
\prog{Kelinci.addCost(diff)} tells the fuzzer to use the cost difference \prog{diff} as cost metric during the input assessment.
We did run the fuzzer for 30 minutes and obtained a maximum cost difference of 47 bytecode instructions, 
with the inputs shown in Listing \ref{lst:results-pwcheck}.

The value of $sec\_2$ is matching the complete value of $pub$, whereas the value of $sec\_1$ is not matching at all.
Note that the fuzzer generated these values on its own, without any further influence by the driver.
The initial input file (the seed file), generated randomly, leads to the cost difference 0.
In fact the difference of 47 instructions is the worst-case scenario and was already retrieved by the fuzzer within 69 seconds.
A value greater than 0 was retrieved by the fuzzer within 5 seconds.

\begin{lstlisting}[caption=Password Checking Driver, label=lst:pwcheck-driver, numbers=left,  stepnumber=1, xleftmargin=1em]
void driver(String[] args) {
	int maxLen = 16;
	int maxData = 3 * maxLen;
	byte[] allBytes = readDataUpToMax(args[0], maxData);
	int len = allBytes.length / 3;
	byte[] pub = Arrays.copyOfRange(allBytes, 0, len);
	byte[] sec_1 =Arrays.copyOfRange(allBytes, len, 2*len);
	byte[] sec_2 = Arrays.copyOfRange(allBytes, 2*len, 3*len);
	
	Mem.clear();
	boolean answer1 = pwcheck_unsafe(pub, sec_1);
	long cost1 = Mem.instrCost;
	
	Mem.clear();
	boolean answer2 = pwcheck_unsafe(pub, sec_2);
	long cost2 = Mem.instrCost;
	
	long diff = Math.abs(cost1 - cost2);
	Kelinci.addCost(diff);
}
\end{lstlisting}

\begin{lstlisting}[caption=Input for Max Cost Difference after 30 min., label=lst:results-pwcheck]
pub=[-48, -4, -48, 7, 17, 0, -24, -48, -48, 16, -48, -3, 108, 72, 32, 0]
sec_1=[72, 77, -16, -66, -48, -48, -48, -48, -28, 0, 100, 0, 0, 0, 0, -48]
sec_2=[-48, -4, -48, 7, 17, 0, -24, -48, -48, 16, -48, -3, 108, 72, 32, 0]
\end{lstlisting}
Afterwards, we used a similar fuzzing driver on the \textit{safe} variant for 30 minutes as well, and we ran \diffuzz again. In this case we have observed no cost differences (i.e., $\delta=0$).
To further check that the program was indeed repaired, we executed the \textit{safe} variant with the inputs obtained with the previous fuzzer run on the \textit{unsafe} variant, obtaining again zero difference.
%: and we obtained zero difference too! This validated that the program was indeed repaired.
%erefore, we conclude that the \textit{unsafe} variant contains a side-channel with regard to the number of instructions performed during program %execution, while the .

\section{Evaluation}
\label{results}
To assess the effectiveness of \diffuzz in identifying side-channel vulnerabilities, we evaluated it on two sets of benchmarks.
%, containing a large number of examples and real world applications.
The first set, taken from \cite{blazer} and \cite{themis}, contains programs with known time and space side-channels, as well as repaired versions.
The second set contains new complex examples from the DARPA Space/Time Analysis for Cybersecurity (STAC) program \cite{stac} as well as popular real-world applications, 
%namely Apache FtpServer~\cite{ftpserver} and AuthMeReloaded~\cite{AuthMeReloaded}, 
on which we identified {\em new} vulnerabilities.
 
For the first set of benchmarks, we compare \diffuzz with \blazer \cite{blazer} and \themis \cite{themis}, two state-of-the-art static analysis tools for detecting side-channel vulnerabilities in \java programs.
\blazer uses decomposition techniques for proving bounded non-interference while \themis uses Quantitative Cartesian Hoare Logic reasoning to check bounded non-interference for \java programs, where the bound $\epsilon$ is set to either 0 or 64 \cite{themis}. 
Since \blazer and \themis are not available, we perform the comparison on the same set of benchmarks that were used to evaluate the respective tools \cite{blazer,themis}.
We received the code for all the benchmarks from the \themis developers. We note that three examples were missing (Apache Shiro, Apache Crypto and bc-java); we therefore could not analyze them.
%These benchmarks contain vulnerable programs as well as repaired versions, that were shown to be safe using the previous tools. 
%We perform the comparison in terms of accuracy and running time
%Our tool and the benchmarks will be made available based on open science principles.
%Please consider our supplemental material with all sources and drivers: \url{https://goo.gl/hUPK1n}
Our tool and the benchmarks are available at our GitHub repository: \url{https://github.com/isstac/diffuzz}

%{\bf Research questions:}
%\begin{itemize}
%\item How does \diffuzz compare with \blazer and \themis on vulnerable and safe versions, %in terms of time and accuracy
%\item Can \diffuzz detect vulnerabilities in complex examples from STAC?
%\item Can \diffuzz detect new vulnerabilities in real-works applications?
%\end{itemize}

\subsection{Experimental Setup}
For each target application, we wrote a driver in \java, following the steps in Procedure~\ref{alg:diffuzzdriver} (Section~\ref{sec:approach}). 
%\afl generates input files, which get parsed by the driver.
%In general, we read all bytes from the input file and distribute the values equally on the public and secret values.
%Then, the target application is instrumented and the fuzzer runs the instrumented application with random inputs and attempts to maximize %the {\em user-defined} function, which computes $\delta$, the cost difference between two executions.
Note that by default, the instrumentor ignores all library code %if not stated explicitly, 
and hence, we specifically copied methods from libraries into the application to instrument them as well.
%For example if a password matching algorithm uses the \method{String.equals} method, we have to create an in-line version of it, which we %place in the application.
%Otherwise the instructions in the \method{String.equals} method, i.e. the crucial part of the matching, would not be counted to the costs.
%\yannic{@Shirin: please doublecheck, I added something on parsing and also on %instrumentation (I removed that we instrument the fuzzer because we skip the main, i.e. %the fuzzer does not get instrumented.}

%Our cost model is the number of bytecode instructions that are executed in the run.
%However, KelinciWCA only is able to measure jumps.
%For \diffuzz purpose, we extended KelinciWCA so that it is able to measure the number of all the instructions. 

Due to the randomness in fuzzing, we run \diffuzz on each application five times, and we report the averaged results. 
All the experiments were performed on a server with \tool{openSUSE Leap 42.3} featuring 8 Quad-Core-AMD 8384 2.7 GHz and 64 GB of memory.
We used \openjdk 1.8.0\_151 and \tool{GCC 4.8.5}. 
Although typically \diffuzz is able to identify a side-channel vulnerability in a few seconds, we run each experiment for 30 minutes. 
%For some applications this time is enough to identify the maximum cost difference, while %in some it only reaches a local maximum. 

As mentioned, \diffuzz reads the inputs to a program from an initial seed file.
In general, we used a randomly generated file.
Some applications get specific types of inputs, such as IBASys that needs an image file.
In that case, we extracted the byte encoding from a random image and used it as the initial input file.
For finding timing side-channels, we use a simple cost model that counts the bytecode instructions during the program run. Both \blazer and \themis similarly count the instructions for their timing side-channel analysis. 
%However, it is difficult to compare precisely the three tools, since they operate on different (intermediate) representations.
%: \diffuzz operates directly on bytecodes while \blazer and \themis operate on Jimple (the intermediate representation of \tool{Soot} \cite{SOOT}). 

\subsection{Evaluating \diffuzz on the \blazer Examples}
% \blazer~\cite{antonopoulos2017decomposition} is a static analyzer that proves non-interference properties. 
We employed \diffuzz on the examples from~\cite{blazer} for evaluating \blazer, which were also analyzed with \themis~\cite{themis}. 
%We obtained these benchmarks from the \themis developers.
%Some of these benchmarks are classic examples from the literature~\cite{genkin2015get,Kocher:1996:TAI:646761.706156,pasareanu2016multi}, %some are micro-benchmarks constructed by the developers of \blazer, and some are extracted from DARPA STAC challenge problems.
They  consist of programs with timing side-channels and
repaired \emph{safe} versions. % that should not exhibit the original vulnerability. 
They are small applications with up to a hundred lines of code.%7 to 106 lines of code. 

Note that the safe version of  \emph{unixlogin} was not executable due to a \prog{NullPointerException} during hash comparison (cf. the Figure 3 in the \blazer paper~\cite{blazer}, second example line 7).
Although \themis did not include this subject, we fixed the issue by adding a dummy comparison of the same MD5 hash of the provided password.% (cf. Listing \ref{lst:unixlogin-safe}). 

%\corina{ say how big the blazer examples are in terms of LOC}

%\begin{lstlisting}[caption= Fixed safe version of \emph{unixlogin}, label=lst:unixlogin-safe, numbers=left,  stepnumber=1, xleftmargin=1em]
%boolean login_safe(String u, String p) {
%        boolean outcome = false;
%        if (map.containsKey(u)) {
%            if (map.get(u).equals(md5(p))) {
%                outcome = true;
%            }
%        } else {
%            boolean unused;
%            String md5_str = md5(p);
%            String md5_str2 = new String(md5_str);
%            if (md5_str.equals(md5_str2)) {
%                unused = false;
%            }
%        }
%        return outcome;
%    }
%\end{lstlisting}

{\tiny
\begin{table*}[ht]
\caption{The results of applying \diffuzz to the \blazer examples. 
%Five experiments were run for each application and the \diffuzz's running time for every experiments is 30 minutes. 
Discrepancies are highlighted in red and italics.}
\centering
%\resizebox{\textwidth}{!}{%
\begin{tabu}{lllllllll}
\hline
\textbf{Benchmark} & \textbf{Version} & \textbf{Average $\delta$} & \textbf{Std. Error} & \textbf{Maximum} &  \multicolumn{3}{c}{\textbf{Time (s)}} \\
& & & & & \diffuzz, $\delta>0$ & \blazer & \themis\\ 
 \hline
  \hline
\textbf{MicroBench} &&&& &  &&\\ \hline
Array & Safe & 1.00 & 0.00 & 1  & 7.40 (+/- 1.21) &1.60 & 0.28 \\
Array & Unsafe & 192.00 & 2.68 & 195 & 7.40 (+/- 0.93) &0.16 & 0.23 \\
\rowfont{\em\color{red}}
%LoopAndbranch & Safe & 277,985,284.00 & 248,637,593.13 & 1,389,926,404 & 115 (+/- 24.80) & 0.23 & 0.33\\
%LoopAndbranch & Unsafe & 2,304,199,806.00 & 841,014,495 & 3,839,675,088  & 160.00 (+/- 34.20) & 0.65 & 0.16\\
LoopAndbranch & Safe & 1,468,212,312.40 & 719,375,479.77 & 4,278,268,702 & 18.60 (+/- 6.40) & 0.23 & 0.33\\
LoopAndbranch & Unsafe & 4,283,404,852.40 & 4,450,278.15 & 4,294,838,782 & 10.60 (+/- 2.62) & 0.65 & 0.16\\
Sanity & Safe & 0.00 & 0.00 & 0 & - & 0.63 & 0.41  \\
Sanity & Unsafe & 4,213,237,198.00 & 60,857,888.00 & 4,290,510,883 & 163 (+/- 40.63)  & 0.30 & 0.17 \\\hline%1,262  (+/- 0.00) 
Straightline  & Safe & 0.00 & 0.00 & 0.00 & - & 0.21 & 0.49\\
Straightline  & Unsafe & 8.00 & 0.00 & 8 & 14.60 (+/- 6.53) & 22.20 & 5.30\\
unixlogin & Safe & 3.00 & 0.00 & 3 & 510 (+/- 91.18) & 0.86 & -  \\
unixlogin & Unsafe & 2,880,000,008.00 & 286,216,701.00 & 3,200,000,008 & 464.20 (+/- 64.61) & 0.77 & - \\  \hline
\textbf{STAC} &&&&&&&\\\hline
modPow1 & Safe & 0.00 & 0.00 & 0 & - & 1.47 & 0.61\\
modPow1 & Unsafe &  2,576.00 & 168.21 & 3,068 & 4.80 (+/- 1.11) & 218.54 & 14.16\\
modPow2 & Safe & 0.00 & 0.00 & 9 & - & 1.62 & 0.75\\
modPow2 & Unsafe & 1,471.00 & 891.00 & 5,206 & 23 (+/- 3.48)  & 7813.68 & 141.36\\
passwordEq & Safe & 0.00  & 0.00 & 0.00 & - & 2.70  & 1.10 \\
passwordEq & Unsafe & 86.40 & 20.31 & 127 & 8.60 (+/-2.11) &  1.30 & 0.39\\ \hline
\textbf{Literature} &&&&&&&\\\hline
k96 & Safe & 0.00 & 0.00 & 0  & -  & 0.70 & 0.61\\
k96 & Unsafe & 338.00 & 185.13 & 3,087,339 & 3.40 (+/- 0.98) & 1.29 & 0.54\\
\rowfont{\em\color{red}}
gpt14 & Safe & 163.20 & 79.84 & 517 & 4.20 (+/- 0.80) & 1.43 & 0.46\\
gpt14 & Unsafe & 6,673,760.00 & 2,211,811.00 & 12,965,890 & 4.40 (+/- 1.03) & 219.30 & 1.25\\
login & Safe & 0.00 & 0.00 & 0 & - & 1.77 & 0.54 \\
login & Unsafe & 62.00 & 0.00 & 62 & 10 (+/- 2.92) & 1.79 & 0.70\\% - 599.80 (+/-199.53)
\hline
\end{tabu}
%}
\label{table:benchmark-blazer}
\end{table*}
}

\subsubsection*{\bf Results}
We summarize the results in Table~\ref{table:benchmark-blazer}. 
The \emph{Average} column shows the (average) cost difference $\delta$ between two executions of an application. 
The \emph{Time} column includes the time that each of the tools (\diffuzz, \blazer, \themis) needed to identify a vulnerability. The numbers for \blazer and \themis are extracted from~\cite{themis}; for the \themis experiments, the bound $\epsilon$ was set to zero. 
For \diffuzz, the time shows the average earliest time that cost difference is bigger than zero, $\delta>0$. The time values for some safe versions are not provided because in those cases the $\delta$ is zero.

The results indicate that \diffuzz is able to accurately identify all the side-channel vulnerabilities in the {\em unsafe} versions. The average cost difference for all \emph{unsafe} programs is more than zero and sometimes it is very large.
%. Interestingly, this value can be very large for some programs such as \emph{sanity}, \emph{loopAndbranch}, \emph{unixlogin}, \emph{gpt14}, and \emph{k96}, e.g., some %hundred thousands, or even millions of bytecode instructions. 
%The running time of \diffuzz is also reasonable, taking an average \mustfix{can we compute an average or just %mention the min and max among all applications? or both average, min and max?} seconds to analyze each benchmark. 

%\subsubsection*{\bf Discrepancies} 
%\pleasenote{Should we remove this heading to be consistent with other sections.}
\diffuzz behaves as expected on the majority of the \emph{safe} versions, finding zero difference, but it also found some discrepancies. 
In two cases (\emph{Array} and \emph{unixlogin}) the differences found (1 and 3) may be attributed to slight discrepancies between the intermediate representations of the different analyses, and can thus be considered negligible. 
However, in two other cases %(\emph{LoopAndbranch} and \emph{gpt14}),
\diffuzz found large $\delta$ values % of 3,839,675,110 and 517, respectively, 
indicating that the repaired versions are in fact \emph{not} safe. 
We discuss them below.
%, although both \blazer and \themis proved them to be safe.  In the following, we provide more details on these two programs. 

\subsubsection*{\bf LoopAndbranch}
%\corina{can we add here the precise values computed by diffuse? otherwise it looks made up}
%\pleasenote{added the values, explained the driver.}
%Listing~\ref{lst:LoopAndbranch} shows the \emph{safe} version of the \emph{Loop\-And\-branch} function. 
%The unsafe version of this function has been repaired so that no matter the value of secret, \emph{taint},
% it executes approximately the same number of instructions, as dictated by the public value \emph{a}. 
Both \blazer and \themis deemed the repaired version of \emph{Loop\-And\-branch} function as \emph{safe}. 
%To analyze this example with \diffuzz, we built a driver where we defined $a\_public$, $taint\_secret1$ 
%and $taint\_secret2$ as integers, and obtained the input values from the first three bytes of the input files.

%{\tiny
%\begin{lstlisting}[caption= \emph{LoopAndbranch} safe version, label=lst:LoopAndbranch, numbers=left,  stepnumber=1, xleftmargin=1em]
% public static boolean loopAndbranch_safe(int a, int taint) {
%        int i = a;
%        if (taint < 0) {
%            while (i > 0) i--;
%        } else {
%            taint = taint + 10;
%            if (taint >= 10) {
%                int j = a;
%                while (j > 0) j--;
%            } else {
%                if (a < 0) {
%                    int k = a;
%                    while (k > 0) k--;
%                }
%            }
%        }
%        return true;
%    }
%\end{lstlisting}
%}

\diffuzz instead identified a huge difference $\delta = 1,389,926,404$ in computed costs, which occurs due to integer overflow.
%generated for the following input values: 
%$a\_public=694963201$,
%$taint\_secret1=2147483647$, 
%$taint\_secret2=1298942977$.
%Upon examining this more closely, we noticed that this side-channel vulnerability occurs due to integer overflow. In particular, the value assigned to $taint\_secret1$ is the maximum integer value in \java. Then at line 7, when 10 is added to it, it exceeds the maximum size of the integer and it becomes a negative value, i.e., -2,147,483,639. As a result, none of the \emph{while} loops is run and the cost remains very small. However, the cost for the other execution with the second secret $taint\_secret2$ is almost twice of the value of the public value. 
%This is because at each iteration in the \emph{while} loop, two bytecode instructions are executed: the check for %the boundary and the subtractive assignment.
In particular, the value assigned by the fuzzer to one of the secrets is the maximum integer value in Java, which gets added to 10, becoming a negative value.
As a result, none of the loops in the code get executed and the cost is very small compared to the cost of the other execution, with the second secret value. 
%As the result, there is a huge difference between the bytecode instructions of two executions. 
This vulnerability, which was confirmed by the developers of \blazer, highlights the importance of handling overflow in analysis tools.

\subsubsection*{\bf gpt14}
%\corina{same here: can you show the actual inputs produced by diffuse?}
%\pleasenote{added the values. explained the driver and constraints}
%We also discuss here another example which is only of a mild concern.
%Listing~\ref{lst:gpt14} shows the safe version of the \emph{gpt14} from \blazer~\cite{blazer}. 
This function computes the modular exponentiation, $a^b mod(p)$, used for the encryption and decryption of messages. 
Here, $a$ and $p$ are public values and $b$ is the secret. 
%which is the remainder when an integer $b$ (the base) raised to the $e^th$ power (the exponent), $b^e$, is divided by a positive integer $p$ (the modulus). 
%In the driver, we obtained the values for $b\_secret1$, $b\_secret2$, $a\_public$ and $p\_public$ by dividing the input files generated by fuzzer into four equal parts. 
%We further added application specific constraints, including: all the values are positive; $b\_secret1$ and $b\_secret2$ have the same bit length; and modulus %is not zero. 

%{\tiny
%\begin{lstlisting}[caption= \emph{gpt14} safe version, label=lst:gpt14, numbers=left,  stepnumber=1, xleftmargin=1em]
%public static BigInteger modular_exponentiation_safe(BigInteger a, BigInteger b, BigInteger p) {
%        BigInteger m = BigInteger.valueOf(1);
%        int n = b.bitLength();
%        for(int i = 0; i < n; i++) {
%            m = m.multiply(m).mod(p);
%            BigInteger t = m.multiply(a).mod(p);
%            if (b.testBit(i)) 
%                m = t;
%         }
%        return a;
%    }
%\end{lstlisting}
%}

\blazer reported this example as safe for a non-zero bound whereas \themis reported it as safe for a zero bound (non-interference). \diffuzz found that even though the repair has substantially reduced the cost difference, still $\delta = 517$ (which is consistent with the \blazer results). 
%The input values generated by \diffuzz which resulted in the maximum difference are as follow:
%{\tiny
%\noindent$b\_secret1 = 374144419156711147060143317158476711873208811257672$\\
%$b\_secret2 = 349732777446035560240497986593709851197853940916256$ \\
%$a\_public = 1724057483480832655435774704206493280677957298307465215$\\
%$p\_public = 95780971304118053647396685922275244456113582954737535$\\
%}
This vulnerability is due to an extra \emph{if} statement that depends on the secret and it was confirmed by the Themis' developers.%$b$.%, i.e., \emph{b.testBit(i)}. 
%In practice, this kind of vulnerability is not of great concern. Still, we consulted with the \themis developers, who indeed confirmed that there is a non-zero %difference for this example. 

\subsection{Evaluating \diffuzz on the \themis Examples}
We further evaluated \diffuzz on the larger \java programs with time and space side-channels from~\cite{themis}.
These programs have up to 20K LOC (although only some smaller parts were analyzed with all 
three tools), and are extracted from complex-real world applications, such as Tomcat, Spring-Security and Eclipse Jetty HTTP web server.
All benchmarks except \emph{DynaTable}, \emph{Advanced\_table}, \emph{OpenMRS} and \emph{OACC} come with a repaired version.
% that does not exhibit the original vulnerability. 
Some of the benchmarks (\emph{Tomcat}, \emph{pac4j}) include interactions with a database.
In our experiments, we created the required databases and run them instead of simulating them with other data structures.
We used the H2 database engine~\cite{h2} to create an SQL database accessible via the JDBC API.
%, and created the necessary tables.
%In the fuzzing drivers we prepare the database, e.g., clearing the database and inserting the appropriate data.
%These benchmarks show that \diffuzz can be employed on different types of applications ranging from web and client-server %applications to those with databases. 
%\corina{how did blazer and themis handle the db? please describe; YN: Blazer had not subject with db, Themis did %not describe it in the paper. I guess that it was not relevant for them, they probably simply wrote a model for %the method which retrieves data.}

%Some of the \themis benchmarks do not include the actual code for some expensive functions 
%(manually created models were used in~\cite{themis} and they were not provided to us).%, which include manually created expensive processing. 
%In addition, \themis works with a Jimple intermediate representation, while \diffuzz works directly with bytecodes, 
%which makes it hard to compare our results exactly. %Although the assumption in \themis is that Jimple code is practically equivalent to bytecodes, in practice this is not necessarily the case.

\subsubsection*{\bf Results} 
Table~\ref{table:benchmark-themis} displays our results; the results for \themis are taken from~\cite{themis}.
Once again, \diffuzz successfully identified vulnerabilities in the unsafe versions of these examples and for the majority of the repaired versions, \diffuzz found only small differences, as expected. 
In one case (jetty), \diffuzz identified a \emph{new} vulnerability in the repaired version.  
Some other examples (Tomcat, pac4j, OACC) also show some discrepancies. We provide more details below.
% that we discussed with the \themis developers.
%The running time of \diffuzz is also quite reasonable, taking an average \mustfix{\emph{??}} seconds to analyze each %benchmark. 
%\corina{we do not discuss time for the blazer benchmarks. we should discuss it in a unified way}
%\subsubsection*{\bf Discrepancies} 
%\themis uses a threshold of $\epsilon = 64$. 
%In the following, we provide more details.

{\tiny
\begin{table*}[ht]
\caption{Comparison against \themis}
\centering
%\resizebox{\textwidth}{!}{%
\begin{tabu}{llllll|ccc}
\hline
\textbf{Benchmark} & \textbf{Version} & \multicolumn{4}{c|}{\textbf{\diffuzz}} & \multicolumn{3}{c}{\textbf{\themis}}\\ 
 &&Average $\delta$ & Std. Error & Maximum & Time (s) $\delta>0$ & $\epsilon=64$ & $\epsilon=0$ & Time (s) \\ 
 \hline
 \hline
Spring-Security & Safe & 1.00& 0.00 & 1 & 9.00 (+/- 1.26) & \cmark & \cmark & 1.70 \\
Spring-Security & Unsafe & 149.00 &  0.00 & 149 & 8.80 (+/- 1.16) &\cmark & \cmark & 1.09 \\
JDK7-MsgDigest & Safe & 1.00 & 0.00 & 1 & 15.80 (+/- 3.93) &  \cmark & \cmark & 1.27 \\
JDK6-MsgDigest & Unsafe & 10,215.00 & 6,120.00 & 34,479 & 7.40 (+/- 1.29) & \cmark & \cmark & 1.33 \\
Picketbox & Safe & 1.00  & 0.00 & 1 & 29.20 (+/-5.00) &\cmark & \xmark & 1.79 \\
Picketbox & Unsafe & 4,954.00 & 1,295  & 8,794 & 16.80 (+/- 2.58) &\cmark & \cmark & 1.55\\
Tomcat & Safe & 12.20 & 1.61 & 14  & 13.80 (+/- 1.29) & \cmark & \xmark & 9.93 \\
\rowfont{\em\color{red}}
Tomcat & Unsafe & 33.20 & 3.40 & 37 & 128.60 (+/- 87.20) & \cmark & \cmark & 8.64  \\
\rowfont{\em\color{red}}
% old version Jetty & Safe &  740.80 & 155.17 & 1,116 & 8.40 (+/- 3.51) & \cmark & \cmark & 2.50 \\
%Jetty & Safe* &  1.80 & 0.18 & 2 & 88.20 (+/- 34.25)  & - & - & - \\
% old version Jetty & Unsafe & 5,038.00 & 4,265 & 24,108 & 9.20 (+/- 1.32) & \cmark & \cmark & 2.07\\
Jetty & Safe & 5454.00 & 1330.88 & 8898 & 9.40 (+/- 1.86) & \cmark & \cmark & 2.50 \\
Jetty & Unsafe & 10786.60 & 2807.51 & 16020 & 7.00 (+/- 1.05) & \cmark & \cmark & 2.07\\
orientdb & Safe & 6.00 & 0.00 & 6 & 3.20 (+/- 0.97) & \cmark & \xmark & 37.99\\
orientdb & Unsafe & 6,604.00 & 3,681 & 19,300 & 3.00 (+/- 0.84) & \cmark & \cmark & 38.09 \\
pac4j & Safe & 10.00 & 0.00 & 10 & 5.00 (+/- 1.22) & \cmark & \xmark & 3.97\\
\rowfont{\em\color{red}}
pac4j & Unsafe & 11.00 & 0.00 & 11 & 8.00 (+/- 2.76) & \cmark & \cmark &1.85 \\
\rowfont{\em\color{red}}
pac4j & Unsafe* & 39.00 & 0.00 & 39 & 10.80 (+/- 5.80) & - & - & - \\
boot-auth & Safe & 5.00 & 0.00 & 5 & 5.20 (+/- 0.20) & \cmark & \xmark & 9.12\\
boot-auth & Unsafe & 101.00 & 0.00 & 101 & 5.20 (+/- 0.20)  & \cmark & \cmark & 8.31 \\
tourPlanner  & Safe & 0.00 & 0.00 & 0 & - & \cmark & \cmark & 22.22\\
tourPlanner  & Unsafe & 522.40 & 18.60 & 576 & 19.20 (+/- 0.80) & \cmark & \cmark  & 22.01\\
DynaTable & Unsafe & 95.80 & 0.44 & 97 & 3.60 (+/- 1.21) & \cmark & \cmark & 1.165 \\
Advanced\_table & Unsafe & 92.40 & 1.54 & 97 & 11.20 (+/- 1.62) & \cmark & \cmark & 2.01 \\
OpenMRS & Unsafe & 206.00 & 0.00 & 206 & 11.60 (+/- 3.22) & \cmark & \cmark & 9.71 \\
\rowfont{\em\color{red}}
OACC & Unsafe & 47.00 & 0.00 & 47 & 7.00 (+/- 1.30) & \cmark & \cmark & 1.83 \\ 
\hline
\end{tabu}
%}
\label{table:benchmark-themis}
\end{table*}
}
\subsubsection*{\bf Jetty}
\themis was used to analyze a known vulnerability in the Eclipse Jetty HTTP web server and a repaired version of it. 
Furthermore, \themis found a similar vulnerability in another part of the Jetty application. 
%Listing~\ref{lst:jetty-safe} shows the code snippet of safe \method{check} method, which was analyzed with \themis.
%Listing~\ref{lst:jetty-safe} shows the string comparison algorithm used by \themis in the safe version of a credential \method{check} method.
%{\bf corina: to be updated:}
% and was provided to us\footnote{Citation [1] in the Themis paper: \url{https://github.com/eclipse/jetty.project/commit/2baa1abe4b1c380a30deacca1ed367466a1a62ea}}. 

%This method matches the user's credential with the expected credential. 
The original unsafe version of the code performs some checking over sensitive credential information by calling the built-in equality method provided by the {\tt java.lang.String library}. 
Since this method returns false as soon as it finds a mismatch between two characters, it introduces a timing side-channel vulnerability. 
%This built-in equality method is replaced with lines 6--9. This repair is very common and has been used in many implementations to avoid time channels. 
The method has been repaired with the one in Listing \ref{lst:jetty-safe}.
This repair is very common and has been used in many implementations to avoid time channels. 

%{\tiny
%\begin{lstlisting}[caption=Jetty safe version analyzed by \themis~\cite{themis}, label=lst:jetty-safe, numbers=left,  stepnumber=1, xleftmargin=1em]
%public boolean check_safe(Object credentials){
%         byte[] digest = null;
%         
%         digest = credentials.toString().getBytes(StandardCharsets.ISO_8859_1);
%         if (digest == null || digest.length != _digest.length) return false;
%         
%         boolean match=true;
%         for (int i = 0; i < digest.length; i++)
%              match&=digest[i] == _digest[i];
%         return match;
%}
%\end{lstlisting}    
%}

{\tiny
\begin{lstlisting}[caption=Jetty safe string comparison analyzed in~\cite{themis}, label=lst:jetty-safe, numbers=left,  stepnumber=1, xleftmargin=1em]
boolean stringEquals(String s1, String s2) {
	boolean result = true;
	int l1 = s1.length();
	int l2 = s2.length();
	if (l1 != l2) result = false;
	int n = (l1 < l2) ? l1 : l2;
	for (int i = 0; i < n; i++)
		result &= s1.charAt(i) == s2.charAt(i);
	return result;
}
\end{lstlisting}    
}

%However, \diffuzz found that this version is still vulnerable with $\delta=1,116$. 
%We created a driver where the secret and public credentials, $secret1$, $secret2$, and $public\_cred$, are defined as \emph{String}s. The driver reads the input file generated by the fuzzer and equally splits and assigns the bytes to the three variables.  
Interestingly, for this example, \diffuzz found that it is still vulnerable with $\delta=5,454$.
% The length of each of variables that generated the maximum cost difference $\delta=1,116$ is 231 characters.\pleasenote{I don't think %we can show the actual values because they are in non-recognizable format. Or perhaps you have a solution for that.}
The reason for this vulnerability is subtle. It turns out that the operation at line 7 is not constant time: it takes either 2 or 3 bytecodes, depending on the outcome of the equality check between the two characters (the operation is optimized for the case that the outcome is false). Although there is a difference of only one bytecode instruction, having this operation in the loop amplifies its impact. This could not be discovered by \themis because in its intermediate representation (Jimple), the operation at line 7 takes constant time.
We further note that we imposed no constraints on the input and this is in line with the \themis experiments. 
The observed difference is proportional with the size of the input, and for small input sizes, both versions could be considered safe. However, for large input sizes, both safe and unsafe versions are in fact not safe. %{\bf say we contacted the developers?}

\subsubsection*{\bf Tomcat, pac4j, OACC}
The vulnerability of \emph{pac4j} is due to the encoding of a password, which is performed during user authentication, 
and is assumed to be expensive.
%is an expensive processing step.
%Depending on where in the code the encoding is performed, an attacker could retrieve whether the given username does exist in the system.
Nevertheless, the code provided by the \themis developers did not include an expensive implementation of the password encoding 
(they instead used a model which was not provided to us).
Since we did not use any models  we could not find a noteworthy cost difference between the provided safe and unsafe versions (cf. Table~\ref{table:benchmark-themis} subject \emph{pac4j Safe} and \emph{pac4j Unsafe}).
We also used another more expensive password encoding method, denoted with a star (*) in Table~\ref{table:benchmark-themis}, which iterates over the password, to get a stronger indication that there is an actual timing side-channel vulnerability (cf. Table~\ref{table:benchmark-themis} subject \emph{pac4j Unsafe*}).
%We further assumed for our experiments that the fuzzed user name does not exceed the length of five characters, and 20 characters for the fuzzed password %respectively.
%Note that the actually password is not relevant here, since the side-channel belongs to the existence of the user name.

We also found vulnerabilities in the unsafe versions of Tomcat and OACC, however the generated $\delta$s were small. Upon consulting with the \themis developers, it appears that, similar to \emph{pac4j}, some manually built models were used, which we could not obtain.
% However, we could not obtain those models to confirm.

\subsection{Employing \diffuzz on New Examples}
We also applied \diffuzz on new \java examples, including two complex applications taken from the Cybersecurity (STAC) program~\cite{stac}: IBASys and CRIME, and two real-world open-source projects: Apache FtpServer~\cite{ftpserver} and AuthMeReloaded~\cite{AuthMeReloaded}. 
%\corina{say the sizes and what we analyzed from them? YN: I added it in the text below, only the numbers for the classes which were involved %in the analysis, is this intuitive?}
%In the following we provide more details about these subjects. 

\subsubsection*{\bf Results} 
Table~\ref{table:benchmark-new} shows the results. % we obtained with \diffuzz on these examples.
For the reported zero-day vulnerabilities, all are confirmed and, collaborating with the developers and the community, solutions have been proposed and at this point most of them have been fixed.
%Here, we explain each of these examples and 
We explain our findings in more details.
{\tiny
\begin{table*}[ht]
\caption{The results of applying \diffuzz on new examples}
\centering
%\resizebox{\columnwidth}{!}{%
\begin{tabular}{lllllc}
\hline
\textbf{Benchmark} & Version & Average $\delta$ & Std. Error & Maximum & Time (s) $\delta>0$\\ 
 \hline
 \hline
\textbf{STAC} & & & & & \\\hline
 CRIME & unsafe & 295.40 & 117.05 & 782 & 7.40 (+/- 1.12) \\
 ibasys (imageMacher) & unsafe & 191 & 20.88 & 262 & 6.20 (+/- 0.66)\\
 \hline
\textbf{Zero-day Vulnerabilities} \\\hline
Apache ftpserver Clear & safe & 1.00 & 0.00 & 1 & 7.20 (+/- 1.24) \\
Apache ftpserver Clear & unsafe & 47.00 & 0.00 & 47 & 6.80 (+/- 1.07) \\
Apache ftpserver MD5 & safe & 1.00 & 0.00 & 1 & 4.20 (+/- 1.93) \\
Apache ftpserver MD5 & unsafe & 151.00 & 0.00 & 151 & 2.80 (+/- 1.11)  \\
Apache ftpserver SaltedPW & (safe) & 176.40 & 6.25 & 198 & 2.20 (+/- 0.73) \\
Apache ftpserver SaltedPW & unsafe & 178.80 & 5.13 & 193 & 3.60 (+/- 1.08)\\
Apache ftpserver SaltedPW* & unsafe & 163.40 & 3.80 & 178 & 5.40 (+/- 0.98) \\
Apache ftpserver StringUtils & safe & 0.00 & 0.00 & 0 &  - \\ 
Apache ftpserver StringUtils & unsafe & 53.00 & 0.00 & 53 & 3.00 (+/- 1.05) \\
AuthMeReloaded & safe & 1.00 & 0.00 & 1 & 7.60 (+/- 0.75) \\
AuthMeReloaded & unsafe & 383.00 & 0.00 & 383 & 9.20 (+/- 1.96)\\
\hline
\end{tabular}
%}
\label{table:benchmark-new}
\end{table*}
}
\subsubsection*{\bf CRIME}
CRIME is an instance of the CRIME attack (``Compression Ratio Info-leak Made Easy'')~\cite{CRIMEattack}, which is as follows. Suppose a user is tricked into visiting a website \verb|attack.com|, which has a malicious script making several requests to \verb|bank.com|. Each request is a concatenation of public input generated by the script and the login (secret) cookie of the user. To avoid latency, protocols such as HTTPS and SPDY compress the requests before they are sent. The communication channel is encrypted, but the adversary can observe the size of the compressed package. When the public input is close to the secret, the compression is more efficient due to the redundancies, and the reduction in the size of the compressed package is more significant. Hence, the adversary can infer information about the secret.
We analyzed the string compression procedure (160 LOC). %, which uses the Lempel-Ziv (LZ77) algorithm. 
It uses various input-output streams and involves complex string manipulations that are difficult to analyze with existing static analysis tools. 

%To analyze CRIME with \diffuzz, we built a driver where the input variables, $cookie\_secret1$, $cookie\_secret2$ and $public\_input$ are defined as byte arrays and are obtained from dividing the input files generated by the fuzzer into three partitions. In the case of CRIME, the cost of a program run is measured in terms of the size of the returned compressed values.
\diffuzz correctly identifies a space side-channel that reveals the secret through the size of the compressed output. 

\subsubsection*{\bf IBASys}
IBASys is a network-based authentication server that uses images in place of textual passwords. To log in, a user supplies a username and a passcode image (\textit{e.g.,} a JPEG image). Following a successful authentication, IBASys replies with a response containing an encrypted session token. This session token could then be used to interact with other services that rely on IBASys for their authentication needs. %Listing~\ref{lst:ibasys} shows the test method that is used for authentication in IBASys. The timing side-channel in this method occurs in %lines 38--48, when the \emph{for} loop gets an early break if the values resulted from some computations on the provided image does not match %the user's stored image. 
We analyzed the authentication procedure (707 LOC), which performs complex image manipulations.% (as only a cropped image is used for authentication). 
%The number of lines of code in the classes that \diffuzz was employed on, i.e., AbstractScalrTest, %ImageMatcherWorker, Scalr, and ScalrApplyTest, is 707. 

%\pleasenote{explained the driver and inputs}
%In the driver, the input variables, $pcode\_secret1$, $pcode\_secret2$ and $image\_public$, are defined as byte arrays, and are extracted from the input files %generated by the fuzzer. The pcode in IBASys has fixed length of 128 bytes, thus when reading from the input file, the first two 128 bytes are assigned to the %first two secrets, and the remaining bytes available in the file are assigned to the image. For the initial seed file, we picked a random image with size of 21 %KB, and two randomly generated byte arrays with length of 128, and wrote them in a file. 
\diffuzz managed to generate input files that are bytecode representations of valid images and it was also able to uncover a timing-channel that is due to early termination in a loop that matches the two (public and private) provided images.
The maximum cost difference found by \diffuzz is $\delta = 262$, where the length of \prog{image\_public} is 18,995 bytes. 

\subsubsection*{\bf Apache FtpServer}
We also applied \diffuzz on the open-source project Apache FtpServer \cite{ftpserver}, which has a very large code base; we focused our analysis on specific classes as reported below.
%In the driver, we encoded the constraints that are specified in the actual classes, e.g. for hash or salt length, and for all experiments with the Apache %FtpServer we applied the constraint that each input value (password or user name) does not exceed the maximum length of 16 characters.

We identified a previously unknown timing side-channel in the class \clazz{ClearTextPasswordEncryptor} (115 LOC), in which the method \method{boolean matches(String, String)} uses the \method{String.equals} method for the comparison of the user provided password and the server-side stored password.
This comparison returns false as soon as a character does not match, and hence, it could be used by a potential attacker to obtain knowledge about the hidden secret password.
We have found this kind of vulnerability also in the classes \clazz{Md5PasswordEncryptor} (185 LOC) and \clazz{SaltedPasswordEncryptor} (211 LOC).
%The results of our analysis are presented in Table \ref{table:benchmark-new}. 
We reported the issues to the developers who confirmed and fixed all of them.
We also analyzed safe versions (provided by the developers) which fixed the issue.
For all of them but one, the safe variant of string comparison did eliminate the vulnerability.
For the class \clazz{SaltedPasswordEncryptor} \diffuzz still detected a vulnerability, so we continued our investigation and discovered that in addition to the matching method the used encryption method leaks information about the generated salt.
We have thus analyzed method \method{String encrypt(String pw, String salt)}, marked with a star (*) in Table \ref{table:benchmark-new}.
We are discussing with the developers with regard to this new vulnerability.

{\tiny
\begin{lstlisting}[caption= Apache FtpServer StringUtils.pad unsafe version, label=lst:padding-unsafe, numbers=left,  stepnumber=1, xleftmargin=1em]
public final static String pad_unsafe(String src, char padChar, boolean rightPad, int totalLength) {
    int srcLength = src.length();
    if (srcLength >= totalLength) return src;
    int padLength = totalLength - srcLength;
    StringBuilder sb = new StringBuilder(padLength);
    for (int i = 0; i < padLength; ++i) {
        sb.append(padChar);
    }
    if (rightPad) {
        return src + sb.toString();
    } else {
        return sb.toString() + src;
    }  }
\end{lstlisting}
}

Note that the salt in \clazz{SaltedPasswordEncryptor} gets randomly generated during encryption. %, i.e. the stored password contains some randomness.
Nevertheless for the \method{matches} method we fuzz the complete stored password including the salt.
%All executions of the fuzzing driver are deterministic.
Furthermore, for the more focused analysis of the \method{encrypt} method we test if the algorithm leaks some information about the used salt via a side-channel.
%Again, we fuzz the secret, i.e. the salt, and hence, each execution of the fuzzing is deterministic.

We have also found a timing side-channel in the method \method{String StringUtils.pad(String, char, boolean, int)} (Listing \ref{lst:padding-unsafe}), which was also confirmed by the developers.
This method leaks the padding in a timing side-channel, from which a potential attacker could obtain the length of the \prog{src} String.
The padding is used to extend a username to fixed length, hence, a potential attacker could obtain the length of a given username, which might be used for further attacks.
The vulnerability is caused by: (1) the early return in line 2, and (2) the for loop in line 5-7, which only runs for \prog{padLength} iterations.
The safe version, provided in our repository %supplemental material, %in Listing \ref{lst:padding-safe} 
solves both issues.

%{\tiny
%\begin{lstlisting}[caption= Apache FtpServer StringUtils.pad safe version, label=lst:padding-safe, numbers=left,  stepnumber=1, %xleftmargin=1em]
%public final static String pad_safe(String src, char padChar, boolean rightPad, int totalLength) {
%    int srcLength = Math.min(src.length(), totalLength);
%    int padLength = totalLength - srcLength;
%    StringBuilder sb = new StringBuilder(padLength);
%    StringBuilder sb_fake = new StringBuilder(srcLength);
%    for (int i = 0; i < totalLength; ++i) {
%        if (i < padLength) {
%            sb.append(padChar);
%        } else {
%            sb_fake.append(padChar);
%        }
%    }
%    if (rightPad) {
%        return src + sb.toString();
%    } else {
%        return sb.toString() + src;
%    }
%}
%\end{lstlisting}
%}

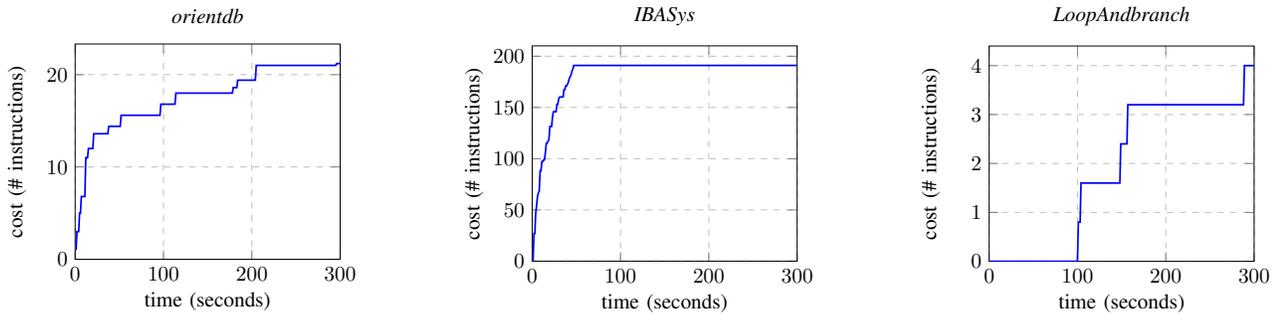
\begin{figure*}[!htb]
\minipage{0.33\textwidth}

\begin{tikzpicture}[scale=\plotscale]
\begin{axis}[
  title=\emph{orientdb},
  xlabel=time (seconds),
  ylabel= cost (\# instructions),
  xmajorgrids=true,  
  ymajorgrids=true,
  grid style=dashed,
  xmin=0, xmax=300,
  ymin=0,
  x label style={at={(axis description cs:0.5,0.03)}},
  y label style={at={(axis description cs:0.06,0.5)}},
  width = \columnwidth,
  legend style={font=\footnotesize,at={(0.64,0.45)},anchor=west},
   ]
\addplot[color=blue,mark=none, thick] table [y=average, x=seconds]{orientdb-unsafe.dat};
\end{axis}
\end{tikzpicture}
%\caption{Themis's \emph{orientdb} unsafe averaged cost over time.}
\label{fig:orientdb_unsafe}

\endminipage\hfill
\minipage{0.33\textwidth}

\begin{tikzpicture}[scale=\plotscale]
\begin{axis}[
  title=\emph{IBASys},
  xlabel=time (seconds),
  ylabel= cost (\# instructions),
  xmajorgrids=true,  
  ymajorgrids=true,
  grid style=dashed,
  xmin=0, xmax=300,
  ymin=0,
  x label style={at={(axis description cs:0.5,0.03)}},
  y label style={at={(axis description cs:0.06,0.5)}},
  width = \columnwidth,
  legend style={font=\footnotesize,at={(0.64,0.45)},anchor=west},
   ]
\addplot[color=blue,mark=none, thick] table [y=average, x=seconds]{ibasys_unsafe.dat};
\end{axis}
\end{tikzpicture}
\label{fig:ibasys_unsafe}

\endminipage\hfill
\minipage{0.33\textwidth}%

\begin{tikzpicture}[scale=\plotscale]
\begin{axis}[
  title=\emph{LoopAndbranch},
  xlabel=time (seconds),
  ylabel= cost (\# instructions),
  xmajorgrids=true,  
  ymajorgrids=true,
  grid style=dashed,
  xmin=0, xmax=300,
  ymin=0,
  x label style={at={(axis description cs:0.5,0.03)}},
  y label style={at={(axis description cs:0.06,0.5)}},
  width = \columnwidth,
  legend style={font=\footnotesize,at={(0.64,0.45)},anchor=west},
   ]
\addplot[color=blue,mark=none, thick] table [y=average, x=seconds]{loop_unsafe.dat};
\end{axis}
\end{tikzpicture}
\label{fig:loop_unsafe}

\endminipage
\caption{Averaged cost over time for \emph{orientdb}, \emph{IBASys}, and \emph{LoopAndbranch} (unsafe versions).}
\label{fig:unsafe_plots}

\end{figure*}

\subsubsection*{\bf AuthMeReloaded}
We have also found an unknown  timing side-channel in the open-source project AuthMeReloaded \cite{AuthMeReloaded}, which is an authentication plugin for Minecraft servers available on GitHub.
It provides features like username spoof protection and anti-bot measures.
%Again, this side-channel is caused by a vulnerable string comparison similar to the one in the Apache FtpServer.
Specifically, we found a vulnerability in the class \clazz{RoyalAuth} (209 LOC), in the inherited method \method{boolean comparePassword(String password, HashedPassword hashedPassword, String name)}.
Similar vulnerabilities have been found in the classes \clazz{Sha256} and \clazz{Pbkdf2}.
The developers fixed these vulnerabilities within a few days by using a constant time comparison algorithm.
%Note that we assumed for all input value that they do not exceed the maximum length of 16 characters.

\subsection{Discussion}\label{subsec:limitations}

One advantage of \diffuzz compared to the other tools is that it not only shows whether an application is vulnerable, but also shows the magnitude of the vulnerability.
This observation can be leveraged to estimate the severity of a vulnerability and it also makes it possible for the developers to compare different repaired versions of an application. 
%For example, if a repaired version decreases the value of cost function several order of %magnitude then it might be enough even though it still has a small side-channel, %unobservable in practice.

\subsubsection*{\bf Analysis Time}
Tables~\ref{table:benchmark-blazer}, \ref{table:benchmark-themis} and \ref{table:benchmark-new} also show the time that each of the tools needed for analysis.
Both \blazer and \themis were run on different hardware, making the timing reported incomparable.
In general, static analysis is shown to be much faster than dynamic analysis; our results show that nonetheless \diffuzz is able to identify vulnerabilities in a reasonable time. In principle \diffuzz can run for a long time and it can still generate new inputs that increase the cost difference.
However, we observed in preliminary test executions that our experiments find a plateau within 30 minutes, which is the time bound we applied.
% for all our experiments.

For our experiments, we observed three different kinds of behavior: (a) \diffuzz identifies a small cost difference very fast, and it increases it over time; (b) \diffuzz identifies a big cost difference very fast and remains in a plateau after a few seconds; and (c) \diffuzz needs a long time to find a cost difference at all.
Cases (a) and (b) were almost equally distributed on our experiments and covered almost all of them.
Only for three experiments we observed case (c).
As an illustration the plots in Figure \ref{fig:unsafe_plots} show the average maximum cost development within the first 5 minutes for the three cases.

%\input{plots/orientdb_unsafe_plot}
%\input{plots/ibasys_unsafe_plot}
%\input{plots/loop_unsafe_plot}
%\input{plots/combined-plots-figure} % moved someway earlier

%Figure \ref{fig:orientdb_unsafe} shows how the computed cost (difference) evolves over time for  \emph{orientdb}, which %represents an example for case (a).
%The average cost value of the five experiments is greater than zero already after the first second, and as shown for the first 5 %minutes of runtime the cost value keeps increasing.
%Figure \ref{fig:ibasys_unsafe} shows the cost development for the \emph{IBASys} subject, which represents an example for case %(b).
%Within a few seconds the cost difference jumps to a big value, where \diffuzz remains on a plateau.
%Figure \ref{fig:loop_unsafe} shows the cost development for  \emph{LoopAndbranch}, which represents an example for case (c).
%\diffuzz needs around 2 minutes to retrieve a cost value greater than zero, which is relatively slow.

%In general the fuzzing time needed to reach a cost difference greater than zero is %difficult to predict because fuzzing, although it is guided to maximize cost, is based on %random generations. 
%, and hence, can be fortunately very fast, or can be unfortunately very slow.
%Nevertheless, our experiments show that even for multiple experiments on the same subject, which mitigates the randomness %problem, there are differences in analysis time.

\emph{Orientdb} (case (a)) checks passwords by comparing between user-given and stored passwords.
The longer the matching prefix is, the higher will be the processing cost. Exact value matching is in general very difficult for fuzzing because it is hard to randomly generate the exact (unlikely) values that match the stored password.
While \diffuzz finds quickly a small prefix, which reveals a cost difference greater than zero, it needs some time to reach a higher value. 

For \emph{IBASys} (case (b)), \diffuzz finds the maximum average value already after a few seconds, and thus leads very fast to the shown plateau value.
The reason could be that the initial seed file guides the fuzzer already into a costly path or that the costly paths have a high probability, and hence, the fuzzer can easily catch them.

For \emph{LoopAndbranch} (case (c)) \diffuzz reaches some parts of the code only with specific values for the secret %(cf. Listing \ref{lst:LoopAndbranch}), 
and this is difficult to achieve with fuzzing.
%It takes more time because generating specific values is very hard for a fuzzing approach, which leverages random mutations.
We believe that the limitations illustrated with cases (a) and (c) can be mitigated by adding further guidance to the fuzzer and by, e.g., combining fuzzing and symbolic execution.
%Symbolic execution is very good at finding inputs that lead to a certain path, and could be used to further push the fuzzer in %regions that are very unlikely to reach by random mutations.

%\subsubsection*{\bf Fuzzing Driver}
%Our approach needs manually created drivers, which trigger the proper executions of the target application. 
%In such a driver, the user needs to specify: how to parse the input file to retrieve valid input values, the entry point to start the target application, and  %how to measure the execution cost. As many applications come with test cases, we
%use them to determine entry points. The drivers can be complex in terms of how to parse data: too many constraints and assumptions on the input data can narrow %down the actual search space while too loose constraints can lead to meaningless results in the analysis. Additionally, it requires some knowledge about the %target application in order to specify good entry points.

\subsubsection*{\bf Vulnerability vs Exploit}
\diffuzz can identify side-channel vulnerabilities but can not assess whether they are exploitable by a real attack.
%The approach presented in this paper focuses only on the identification of a side-channel %vulnerability.
The synthesis of a real attack, which would be necessary to assess the severity of the found vulnerability, is out of scope for this work.
Nevertheless, we believe that our contribution is a first step in this direction.

%\shirin{I also added the CRIME and Compress method. They are also big. Discuss if we want them to be in the paper.} 
%
%\begin{lstlisting}[caption=CRIME, label=lst:crime, numbers=left,  stepnumber=1, xleftmargin=1em]
%public class CRIME
%{
%    private static final String IV = "AAAAAAAAAAAAAAAA";
%    private static final String HELP = "CRIME send <filename> <comment>\n";
%    
%    public static void send(final String filename, final FileOutputStream fos, final Cipher c, final String comment) {
%        final CipherOutputStream cos = new CipherOutputStream((OutputStream)fos, c);
%        final byte[] commentBytes = comment.getBytes();
%        try {
%            final byte[] bytes = Files.readAllBytes(Paths.get(filename, new String[0]));
%            final byte[] all = Arrays.copyOf(bytes, bytes.length + commentBytes.length);
%            System.arraycopy((Object)commentBytes, 0, (Object)all, bytes.length, commentBytes.length);
%            final byte[] compressed = LZ77T.compress(all);
%            cos.write(compressed);
%        }
%        catch (IOException e) {
%            throw new RuntimeException((Throwable)e);
%        }
%    }
%}
%\end{lstlisting}
%

\section{Related Work}
\label{SEC:related}

%\begin{itemize}
%\item Checking non-interference and its variations
%\item \themis and \blazer: what they do exactly
%\item side-channel analysis
%\item Differential Fuzzing, Fuzzing
%\item Resource analysis: in particular Badger
%\end{itemize}

\diffuzz is related to a large body of work on checking non-inter\-fer\-ence via self-com\-po\-si\-tion~\cite{Barthe04}. 
%B. Almeida, M. Barbosa, J. S. Pinto, and B. Vieira. Formal verification of side-channel countermeasures using self-
%composition. Science of Computer Programming, 78(7),2013.
For instance, related work~\cite{SelfComp} presents a self-composition approach to timing-channel analysis, which however does not check bounded non-interference.
%Instead of checking non-interference%, which might not hold for 
%most realistic applications, 
%we check bounded non-interference, which 
%tolerates small differences in cost between different secret-dependent paths.
% that are too small to be observable. 
We already compared with the most recent related tools,  \blazer~\cite{blazer} and \themis~\cite{themis}.
%, with which we already compared \diffuzz.
%We already compared \diffuzz with them throughout the paper.
%The main difference is that these previous tools perform static analysis while \diffuzz performs 
%dynamic analysis. In principle static analysis can provide {\em guarantees} w.r.t. 
%absence of side-channels but we have shown examples that in practice these guarantees do not hold.
%Furthermore they can produce false alarms. \diffuzz does not give false alarms and also produces 
%input values that can be used to reproduce and fix the found vulnerabilities. 

CoCoChannel~\cite{CoCoChannel} uses static analysis for finding 
side-channel vulnerabilities and presents a comparison with \themis and \blazer on the same benchmarks, showing better scalability.
While CoCoChannel also found discrepancies in the \themis and \blazer benchmarks, 
the approach still fails to report vulnerabilities for the repaired versions in, e.g., loopAndBranch and jetty.%, due to the same reasons.
%i.e., the intermediate representation may be inaccurate and the analysis does not handle overflow.

Stacco~\cite{Stacco} also uses a  differential analysis for finding timing side-channels, using random inputs.
However, Stacco does not perform directed fuzzing, it does not check bounded non-interference, and it does not address Java. 
%does not direct the generation of inputs to expose differences in executions. 
%This is necessary as in general, it is not the case that any two random inputs will expose a side-channel.

There is a large amount of related work on side-channel analysis, for example
\cite{Kocher:1996:TAI:646761.706156,DBLP:conf/ches/AgrawalRR03,Brumley:2003:RTA:1251353.1251354,Kopf:2007:IMA:1315245.1315282,Chen:2010:SLW:1849417.1849974,DBLP:conf/sp/MardzielAHC14,Do2015}.
The most successful approaches use abstract interpretation (for cache side-channels analysis) \cite{Kopf:2012:AQC:2362216.2362268,Doychev:2013:CTS:2534766.2534804,DBLP:conf/essos/MantelWK17} and are thus quite different than \diffuzz. 
%Our approach differs from these works by using fuzzing and differential analysis. 
Other techniques~\cite{pasareanu2016multi,DBLP:conf/csfw/PhanBPMB17,Bang:2016:SAS:2950290.2950362} use symbolic execution and constraint solving with model counting for quantifying side-channel leakage and for synthesis of attacks.
They address \java programs, but may have scalability issues, due to the expensive constraint manipulation.

Other related techniques aim to quantify leakage using Monte Carlo sampling \cite{Chothia:2014:LEI:3088411.3088424,DBLP:conf/fm/KawamotoBL16}. 
In contrast to \diffuzz, these techniques provide {\em quantitative} results, but they may be imprecise in practice.

%Fuzz testing tools, such as \afl~\cite{afl}, have been very successful at finding bugs and vulnerabilities in a variety of applications, ranging from image %processors and web browsers to system libraries and various language interpreters. For example, \afl was instrumental in finding several of the Stagefright %vulnerabilities in Android, the Shellshock related vulnerabilities CVE-2014-6277 and CVE-2014-6278, 
%vulnerabilities in \tool{BIND} (CVE-2015-5722 and CVE-2015-5477), 
%as well as numerous bugs in (security-critical) 
%popular applications and libraries such as \tool{OpenSSL}, \tool{OpenSSH}, \tool{GnuTLS}, \tool{GnuPG}, \tool{PHP}, \tool{Apache}, and \tool{IJG jpeg}. 
%%%\tool{libjpeg-turbo} and many more. Fuzzers use heuristic algorithms to mutate user-provided inputs to increase coverage, with the goal of finding crashes %and other vulnerabilities. In contrast \diffuzz uses fuzzing to perform a relational analysis, where the goal is to maximize the difference in resource usage %for two programs.

%Differential fuzzing has found applications in the security domain, e.g.,~\cite{DBLP:conf/sp/PetsiosTSKJ17}. Differential fuzzing uses similar (but not %identical) programs as cross-referencing oracles to find semantic bugs that do not exhibit explicit error behavior like crashes or assert violations. In %contrast we define here a novel application of differential fuzzing, which works by analyzing two copies of the same program with the goal of finding side-%channel vulnerabilities.

Fuzzing has received renewed interest in the software engineering community, with many recent approaches reported~\cite{PerfFuzz,Singularity,FairFuzz,Badger}.
Most related are techniques that use fuzzing alone~\cite{PerfFuzz,Singularity} or a combination of fuzzing and symbolic execution~\cite{Badger}
to analyze the algorithmic complexity of programs, by monitoring a resource consumption. In particular, Badger~\cite{Badger} also uses Kelinci and AFL for the fuzzing part.
None of these works address side-channel analysis.
%aim to maximize the difference in resource usage, as required by side-channel analysis.
%Our work differs in that it analyzes {\em two copies} of the program and rather than trying to maximize the resource (time or memory) usage of the program, 
%\diffuzz aims to maximize the {\em difference} between the resource usage in the two copies of the program.

\section{Conclusions and Future Work}
We presented \diffuzz, the first differential fuzzing approach for automatically finding side-channel vulnerabilities.
We have shown that \diffuzz can keep up with existing approaches such as \blazer and \themis.
Furthermore, \diffuzz found new vulnerabilities in popular open-source \java applications such as Apache FtpServer.
%In our future work we plan to investigate combining differential fuzzing with symbolic %execution to guide the fuzzer faster to high cost differences between two program %executions.
%As discussed, these vulnerabilities may only show weaknesses in the code, but they are not necessarily exploitable in practice.
%As a next step we plan to investigate the use of fuzzing to automatically synthesize attacks to check that the vulnerabilities are exploitable.
In the future, we plan to explore automated repair methods to eliminate the vulnerabilities discovered with \diffuzz.
Additionally, we plan to augment our work with statistical guarantees similar to the STADS framework \cite{stads}.

\section*{Acknowledgment}
This material is based on research sponsored by DARPA under agreement number FA8750-15-2-0087.
The U.S. Government is authorized to reproduce and distribute reprints for Governmental purposes notwithstanding any copyright notation thereon.
This work is also supported by the German Research Foundation (GR 3634/4-1 EMPRESS).

\balance
\bibliographystyle{plain}
\bibliography{refs,icse18,paper,issta18,ase18,fse18}

\end{document}